\documentclass{article}
\usepackage{spconf,amsmath,graphicx}
\usepackage{amssymb}
\usepackage{bm}
\usepackage{color}
 
\def\x{{\bm x}}

\def\R{{\mathbb R}}
 \def\N{{\mathcal N}}
 \def\H{{\mathcal H}}
 \def\ker{{\kappa}}
 \newcommand*{\QED}{\hfill\ensuremath{\blacksquare}}%

 \newtheorem{mylem}{Lemma}
 \newtheorem{myrem}{Remark}
 \newtheorem{mytheorem}{Theorem}
 
\title{A Diffusion  Kernel LMS algorithm for nonlinear adaptive
networks}
\name{Symeon Chouvardas,  Moez Draief}
 
\address{Mathematical and Algorithmic Sciences Lab,\\
Huawei France R\&D,\\
Paris, France.\\
{symeon.chouvardas,moez.draief}@huawei.com
}
\begin{document}
%
\maketitle
\begin{abstract}
This work presents a 
distributed algorithm
for nonlinear adaptive learning. 
In particular,  a set of nodes obtain measurements, sequentially one per time step, which are related via a nonlinear function; their goal is to   
collectively minimize a cost function by employing a diffusion based Kernel Least Mean Squares (KLMS).
The algorithm follows the  Adapt Then Combine
mode of cooperation. Moreover, 
the theoretical properties of the algorithm are 
studied and it is proved that under certain assumptions
the algorithm suffers a no regret bound. Finally,  
comparative experiments verify that the proposed scheme outperforms
other variants of the LMS. 
 
\end{abstract}
\vspace{-0.5cm}
\section{Introduction}
\label{sec:intro}
In recent years, the interest in the 
topic of \textit{distributed learning} and inference
has grown rapidly.
This is mainly due to the constantly
increasing requirements for memory 
and computational resources,
demanded by modern applications, so as to cope with 
the  huge amount 
of available data.
 {These data ``spring" from several sources/applications, such as
communication, imaging, medical
platforms as well as social-networking sites, e.g., \cite{SlGiMa14}.} 
A natural way,
to deal with  the large number of data, 
which need to be processed,
is to split the problem 
into subproblems and resort to 
distributed operations \cite{ChSaLi07,zikopoulos2011understanding}. 
Thus, the development of algorithms dealing with 
such scenarios, where
the data are not  available in a single location 
but are instead spread out over
multiple locations,   becomes essential. 

An important application within the distributed
learning context is the one of  \textit{distributed adaptive learning}, \cite{Sa13}. In a nutshell, this problem considers a decentralized 
network {consisting of}  nodes 
 interested in performing 
a specific task, which can be, for instance,
parameter estimation, classification, etc.
The nodes constantly obtain 
new measurements and  they continuously adapt and learn;
this gives them the capability to track
and adapt to changes in the environment.  On top of that,
it is assumed that there is no central node which could
perform all the necessary operations and, so,
 the nodes act as independent learners 
 and perform the computations by themselves.
 Finally, the task of interest
  is considered to be common or similar across the nodes 
  and, to that direction, they cooperate with each other.
 Cooperation has been proved to be beneficial
 to the learning process since it improves
 the learning performance, \cite{Sa13}.

This paper is concerned with the problem
of distributed adaptive learning in Reproducing
Kernel Hilbert spaces (RHKS).
To be more specific, we consider an ad--hoc
network the nodes of which obtain input/output
 measurements,
sequentially, one per time step, related via a
\textit{nonlinear} unknown function. 
To cope with this nonlinearity we resort to the family of
the kernel--based algorithms for nonlinear 
adaptive filtering. In particular, the proposed algorithm
belongs to the Kernel LMS (KLMS) algorithmic
family and follows the diffusion rationale
for cooperation among the nodes.

\textbf{Related Work:} Several studies for distributed 
adaptive estimation of linear systems have been proposed in the literature.
 These include diffusion based algorithms, e.g., \cite{ChSlTh12,LoSa08,CaSa10},  
  consensus ones, e.g., \cite{ScMaGi09,MaScGi09}, as well 
as algorithms for multitask learning \cite{ChCeSa14,PlBoBe15}.
The problem of non--linear adaptive estimation in RKHS has been studied,
e.g., \cite{bouboulis2011extension,bouboulis2012augmented,slavakis2008online,slavakis2009adaptive}.
A recent study, which considers the problem of nonlinear adaptive filtering
in distributed networks, can be found in \cite{DKLMSrichards}.
The major differences of this paper with our work
are summarized in the sequel. 
First, the authors
consider a  {predefined dictionary},
which essentially makes the dimension of the
problem finite and equal to the number of elements
of the dictionary.
On the contrary, here, we consider the
general case, where the  dictionary is allowed to grow
as time increases, and we present a more general
form of the algorithm.
Furthermore, here, we study for the first time
the theoretical properties of the Diffusion Kernel LMS (DKLMS)
and we derive regret bounds for the proposed scheme.

\textbf{Contributions:} In this paper,
we propose a novel nonlinear distributed algorithm
for adaptive learning. In particular, we propose
a KLMS  algorithm, which follows
the diffusion rationale. The Adapt Then Combine
 mode of cooperation
among the nodes is followed.
To be more specific, we assume that the nodes obtain measurements,
which arrive sequentially and are related
via a nonlinear system. The goal
is the minimization of the expected value
of the networkwise discrepancy between the 
desired output and the estimated one.
To that direction, at each step, the nodes:
a) perform a local update step exploiting their most recent
measurements, b) cooperate with each other,
in order to enhance their estimates. 
Comparative experiments  illustrate that the proposed scheme
outperforms other LMS variants
and the theoretical properties of the proposed
scheme are discussed.

\emph{Notation:}
Lowercase and uppercase boldfaced letters stand for
vectors and matrices respectively.  The symbol
  $\mathbb{R}$ stands for the set of real numbers
and $\mathbb{N}$ for the set of nonnegative integers.
$\mathcal{H}$ denotes an infinite dimensional
Hilbert space equipped with an inner product denoted by
$\langle f_1,f_2\rangle, \ \forall f_1, \ f_2\in \H$;
the induced norm  is given by $\Vert f \Vert = \sqrt{\langle f,f\rangle}$.
Given a set $\mathcal{S}$, with the term $\vert\mathcal{S}\vert$
we denote its cardinality.
 
\vspace{-0.5cm}
\section{Problem Statement}
\label{sec:format}

\begin{figure}[tb]
\centering
\includegraphics[width=2in]{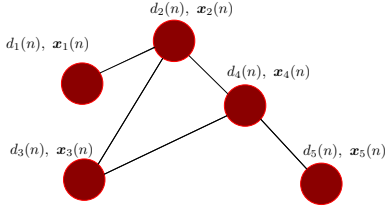}
\caption{An ad--hoc Network}
\label{fig:adhoc}
\end{figure}

We consider an ad--hoc network  illustrated in Fig. \ref{fig:adhoc}, consisting
of
$K$ nodes.  Each node, $k\in\mathcal{N}:=\{1,\ldots,K\}$, at each discrete time instance $n\in\mathbb{N}$, has access to a scalar $d_{k}(n)\in\mathbb{R}$ and a vector $\x_{k}(n)\in\R^m$, which are related via 
\begin{equation} d_{k}(n)=f^o(\x_{k}(n))+v_{k}(n), \label{eqnl} 
\end{equation} 
where  {$f^o: \ \R^m\mapsto \R$} is an unknown yet common to all the nodes
function belonging to the Hilbert space $\mathcal{H}$
and the term $v_{k}(n)$ stands for the additive noise process.
The overall goal is the estimation of a function, $f$, which minimizes the cost:
\begin{equation}
J(f) = \sum_{k\in\N} \mathbb{E}\lbrace \left(d_{k}(n)-f(\x_{k}(n))\right)^2 \rbrace,
\label{eqnlcostd}
\end{equation} 
in a \textit{distributed} and \textit{collaborative} fashion{; that is the nodes want to minimize the cost \eqref{eqnlcostd} by relying
solely on local processing as well as interactions with their neighbors}.
\vspace{-0.3cm}
\subsection{Linear Diffusion LMS}
\label{sec:DLMS}
In order to help the reader grasp the concept
of the diffusion LMS, in this section we  describe
the linear scenario, i.e., the one where the function to be estimated
is a vector, say $\bm{w}_*\in\R^m$, and  \eqref{eqnl} essentially becomes:
\begin{equation} d_{k}(n)=\bm{w}_*^T\x_{k}(n)+v_{k}(n). \label{eql} 
\end{equation} 
The cost function to be minimized in that case can be written 
as follows:
\begin{equation}
J(\bm{w}) = \sum_{k\in\N} \mathbb{E}\lbrace \left(d_{k}(n)-\bm{w}^T\x_{k}(n)\right)^2 \rbrace.
\label{eqlcost} 
\end{equation} 
The cost  \eqref{eqlcost} includes information coming from the whole network 
and in order to minimize it, global knowledge is required.
Nevertheless, in  distributed and decentralized learning
 each node can only interact and exchange information with 
its neighborhood which will be denoted by $\mathcal{N}_k, \ \forall k\in\N$.  
A fully distributed algorithm, which can be employed for the 
estimation of $\bm{w}_*$  
is the diffusion LMS (see for example \cite{Sa13,CaSa10}). 
The starting point of this scheme is a modification of the  steepest--descent method, 
which is properly reformulated so as to enable distributed operations and to avoid
any global computation (the interested reader is referred to \cite{Sa13}).
In addition, the instantaneous approximation is adopted,
according to which the statistical values are substituted
by their instantaneous ones, e.g., \cite{Ha08}. Each node
$k\in\N$ updates the estimate $\bm{w}_k(n)$ at each time step according to the following iterative scheme:
\begin{align}
\bm{w}'_k(n) &= \bm{w}_k(n-1) + \mu_k e_k(n)\x_k(n)\label{eqLMS1} \\
\bm{w}_k(n) &= \sum_{l\in\N_k} a_{k,l} \bm{w}'_l(n)\label{eqLMS2},
\end{align}
where $e_k(n):=d_k(n)-\bm{w}_k^T(n-1)\x_k(n)$
 and $\mu_k$ is the step size. Furthermore,  $a_{k,l}$ stand for
combination coefficients, which have the following properties
$ a_{k,l} = 0, \ l\notin \N_k, \ a_{k,l} > 0, \ l\in \N_k,
 \sum_{l\in\N} a_{k,l} = 1.$
  A common choice, among others, for choosing these coefficients
 is the Metropolis rule, in which the weights equal to:
 \begin{equation*} 
 \footnotesize
 a_{k,l}= \begin{cases} \frac{1}{\max\lbrace\vert\N_k\vert,\vert\N_l\vert\rbrace}, \ &\mathrm{if} \ l\in\N_k, \ \mathrm{and} \ l\neq k\\ 1- \sum_{l\in\N_k \setminus {k}} a_{k,l}, \ &\mathrm{if} \ l=k\\ 0, &\ \mathrm{otherwise}. \end{cases} 
 \end{equation*}
 The intuition behind  the scheme presented in \eqref{eqLMS1},
 \eqref{eqLMS2}, can be summarized as follows.
  In the first step, node $k$ updates its estimate 
  using an LMS based update (adaptation step) exploiting
  local information.
  In the sequel,   $k$ cooperates with its neighborhood by combining their intermediate estimates to obtain its updated  estimate $\bm{w}_k(n)$.
 The
 weights $a_{k,l}$ assign a non--negative weight to the estimates
 received by the neighborhood, whereas they are equal to zero
 for the rest of the nodes. 
 Hence, each node \textit{aggregates} the  information
 received by the neighborhood.  
  This scheme is also known as Adapt Then Combine (ATC)  diffusion strategy.
  \vspace{-0.3cm}
\subsection{Centralized Kernel LMS}
\label{sec:KLMS}
\subsubsection{Preliminaries}
Now, let us provide with a few elementary properties of the RKHS,
which will be used in the sequel.  {Throughout  this section the node subscript will be suppressed since we will describe  
properties of centralized learning.}
We consider a real Hilbert space $\H$ comprising functions
defined on $\R^m$; that is $f: \R^m\mapsto \R$.
The function 
$\kappa: \ \R^m\times \R^m \mapsto \R$  
will be called a reproducing kernel of $\H$ if the following
properties hold:
\begin{itemize}
\item $\forall \x\in\R^m$ the function $\ker(\x,\cdot)$
belongs to $\H$.
\item  $\forall \x\in\R^m, \ f\in\H$, it holds
that $f(\x)=\langle f, \ker(\x,\cdot)\rangle$.
\end{itemize}
If these properties hold then $\H$ is called a Reproducing Kernel Hilbert Space \cite{SmScBe98,Th15}.
A typical 
example is the Gaussian kernel 
with definition:
$\ker(\x_i,\x_j) = \exp(-\beta\Vert \x_i-\x_j \Vert^2), \ \beta>0.$
A very important property, which will be exploited in the sequel states that points in the RKHS can be written as follows:
\begin{equation}
f = \sum_{n=0}^{\infty} \alpha_n \ker(\x(n),\cdot),
\label{eqkertr}
\end{equation}
where $\alpha_n \subset \R$.
Finally,  the reproducing kernel is continuous, symmetric  and positive-definite. 
\vspace{-0.51cm}
\subsubsection{Kernel LMS}
Kernel LMS, which was originally proposed in \cite{LiPoPuPr08},
is a generalization of the original LMS algorithm, which utilizes
the transformed input, i.e., $\ker(\x(n),\cdot)$, at each iteration step. Put in mathematical terms, the recursion of the KLMS is given by:
\begin{align}
e(n) &= d(n) - \langle f_{n-1}, \ker(\x(n),\cdot) \rangle\label{eqKLMS1}\\
f_n &= f_{n-1} + \mu e(n) \ker(\x(n),\cdot).\label{eqKLMS2}
\end{align}
Since the space $\H$ may be infinite dimensional, it may be
difficult to have direct access to the transformed input data
and the function $f_n$. However, if we go back to  \eqref{eqnl}
and forget the distributed aspect for now,
we can see that  the quantity of interest is $f(\x(n))$,
which can be computed exploiting \eqref{eqkertr}.
In particular, following similar steps as in \cite{LiPoPuPr08}
it can be shown that the KLMS recursion can be equivalently written:
\begin{align}
e(n) &=  d(n) -  \mu \sum_{i=1}^{n-1} e(i)\ker(\x(n),\x(i))  \\
f_{n}(\x(n)) &= \mu \sum_{i=1}^n e(i)\ker(\x(n),\x(i)). \label{eqKLMS3}
\end{align}
Note that this reformulation is very convenient as 
 it computes the response of the estimated
 function to the input, without any need to estimate the function itself. 
 
  \vspace{-0.4cm}
\section{The Diffusion Kernel LMS}
In this section we describe the proposed algorithm
together with its theoretical properties.
Recall the problem under consideration, discussed in Section \ref{sec:format}.
 Our goal, here, is to bring together the tools
described in Sections  \ref{sec:DLMS}, \ref{sec:KLMS}  
 and derive a Kernel based LMS algorithm
suitable for distributed operation.
Our starting point will be the  ATC--LMS described in
\eqref{eqLMS1}-\eqref{eqLMS2} and we will employ the non--linear 
transformation on the input (similarly to \eqref{eqKLMS1}, \eqref{eqKLMS2}). The resulting recursion $\forall k\in\N$ is:
\begin{align}
f'_{k,n}   &= f_{k,n-1}   + \mu_k e_k(n)\ker(\x_k(n),\cdot)\label{eqDKLMS1} \\
f_{k,n}&= \sum_{l\in\N_k} a_{k,l} f'_{l,n}\label{eqDKLMS2},
\end{align}
{where $e_k(n)$ is defined similarly to \eqref{eqKLMS1}.}
Despite the fact that this seems a trivial generalization of \eqref{eqLMS1}, \eqref{eqLMS2},
as we have already discussed previously, one cannot resort directly to this form of iterations, since access to the transformed data 
may not be possible.

Exploiting the lemma, which will be presented shortly, we can
 bypass the aforementioned problem,
 by deriving  the inner product between the obtained function and the transformed input 
vector in a closed form.
Before we proceed, let us introduce some notation.
The networkwise function at time $n$ is denoted by
$\underline{f}_n:=[f_{1,n},\ldots,f_{K,n}]^T\in\bm{\H}$,
where the Cartesian product $\bm{\H}:=\underbrace{\H\times \ldots \times\H}_{K \ \mathrm{times}}$. Similarly, we define the networkwise 
input: ${\ker}(\underline{\x}(n),\cdot)=[\ker(\x_1(n),\cdot),\ldots,\linebreak\ker(\x_K(n),\cdot)]^T\in\bm{\H}$ and  $\bm{g}(n) = [\mu_1 e_1(n),\ldots,\mu_Ke_K(n)]^T\in\R^m$. Finally, we gather the combination coefficients to the $K\times K$ matrix $\bm{A}$, the $k,l$--th entry of
which contains  $a_{k,l}$.
It can be readily shown that \eqref{eqDKLMS1}-\eqref{eqDKLMS2}
can be written for the whole network in the following
compact form: 
\begin{equation}
\underline{f}_n =\bm{A}\left(\underline{f}_{n-1}+\bm{g}(n){\ker}(\underline{\x}(n),\cdot)\right).\label{eqnetDKLMS}
\end{equation}
\begin{mylem}
Assume that $f_{k,0}=0,\ \forall k\in\N$.
Then equation \eqref{eqnetDKLMS} can be equivalently
written:
\begin{equation}
\underline{f}_n =\sum_{i=1}^n \bm{A}^{n-i+1}\bm{g}(i){\ker}(\underline{\x}(i),\cdot). 
\label{eqnetDKLMS2}
\end{equation}
Hence, the vector of responses, $\tilde{\bm{d}}(n):=[\tilde{d}_1(n),\ldots,\tilde{d}_K(n)]^T$ ,  at time instance $n$, is given by
$\tilde{\bm{d}}(n)=\underline{f}_n({\ker}(\underline{\x}(n),\cdot))
=\sum_{i=1}^n \bm{A}^{n-i+1}\bm{g}(i){\ker}(\underline{\x}(i),\underline{\x}(n))$\QED
\end{mylem}
The proof, which follows mathematical induction,
 is omitted due to lack of space and will be presented elsewhere.
 \begin{myrem}
 \textbf{Coefficient  Reduction over Time:} If we take a closer look on \eqref{eqnetDKLMS2}
 it can be seen that the number of coefficients
 one has to store as well as the required number of operations
 grow  as time evolves.
 Several sophisticated techniques, 
   which set most of the
 coefficients to zero while  {avoiding} performance degradation, 
 have been proposed in the literature, e.g.,
 \cite{GaChRiHu14}. As it will become apparent in the simulations section,
 here we adopt a simple  method; that is, we apply a buffer of
 size $L$. In that case, we store the  $L$ most recent coefficients
 and   \eqref{eqnetDKLMS2} becomes $\underline{f}_n =\sum_{i=\max(1,n-L+1)}^n \bm{A}^{n-i+1}\bm{g}(i){\ker}(\underline{\x}(i),\cdot).$  
 \end{myrem}
 \begin{myrem}
 \textbf{Coefficient  Reduction over Space:} 
 It can be shown (see \cite[Appendix E]{Sa13}) that the $k,l$--th
 entry of the $i$--th power  of the matrix $\bm{A}$
 equals to: \linebreak$[\bm{A}^i]_{kl} = \sum_{j_1}^K\sum_{j_2}^K\ldots\sum_{j_{i-1}}^K a_{kj_1}a_{kj_2}\ldots a_{j_{i-1} l}$.
 From the last relation it is not difficult to obtain
 that the node $k$ exploits information from nodes which do not belong to its neighorhood. 
 However, this does not break the rules of decentralized learning, since it also holds that $[\bm{A}^i]_{kl}$ will be nonzero
 iff the distance, measured in hops, between $k,\ l$ is smaller or equal than $i$ hops. Hence, the nodes can send their input vectors to their neighbors,
 which in turn will forward them to their neighbors and so on.
 This increases the network load and one can avoid it
 by setting some of the weights to zero, as discussed on
 Remark 1. A simple strategy is to set to zero all
 the coefficients that belong to nodes which don't
 belong to the neighborhood.
 \end{myrem}
 \subsection{Theoretical Properties}
 In the sequel, we will present the regret 
 bound of the proposed scheme and in particular
 we will show that this grows sublinearly 
 with the time.
\begin{mytheorem}
Under certain assumptions about the boundness of the input, the step-size
and the combination weights, the networkwise regret is bounded by
\begin{equation}
\sum_{i=1}^N\sum_{k\in\N}(J_{k,i}(f_{k,i-1})-J_{k,i}(g)) \leq \gamma\sqrt{N} +\delta, \forall g\in\H
\end{equation}
where  $J_{k,n}(f)=\frac{1}{2}(d_{k}(n)-f(\x_k(n)))^2$ and $\gamma, \ \delta$ are positive constants. \QED
\end{mytheorem}
\textbf{Proof:}
The proof is omitted due to lack of space
and will be presented elsewhere.
\vspace{-0.5cm}
\section{Simulations}
\begin{figure}[tb]
\centering
\includegraphics[width=2.3in]{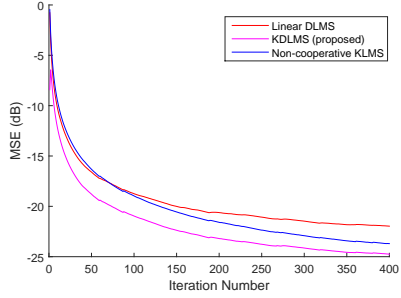}
\caption{Average MSE for the first experiment}
\label{fig:exp1}
\end{figure}
\begin{figure}[h]
\centering
\includegraphics[width=2.3in]{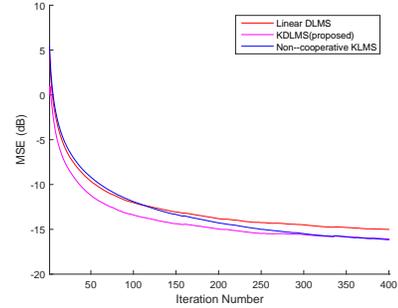}
\caption{Average MSE for the second experiment}
\label{fig:exp2}
\end{figure}
In this section, the performance of the proposed algorithm
is validated within the
distributed nonlinear adaptive filtering framework.
We consider a network comprising $K=10$ nodes
and  a distributed version of the problem studied
in \cite{NaPa90,Ma04},  for which the input and the output are related via:
\begin{align*}
\footnotesize
 y_k(n) = \frac{ y_k(n-1)}{1+ y_k^2(n-1)} +x^3_k(n),\  d_k(n) &=  y_k(n) +v_k(n),
 \end{align*}
 where $v_k(n)$ is Gaussian with variance $10^{-3}$
and the input $x_k(n)$ is also Gaussian with variance
$0.1 \chi_k$, where $\chi_k\in[0.5,1], \forall k\in\N$
with respect to the Uniform distribution.  
We compare the proposed algorithm with: a) the linear diffusion 
 LMS, b)  the non--cooperative KLMS , i.e., the
KLMS where the nodes do not cooperate with each other.
For the Kernel based algorithms we employ the Gaussian Kernel
with $\alpha=1.1$
 and we choose a step--size equal to $\mu=0.6$ 
for all the algorithms.
Furthermore, the combination weights are chosen with respect to the Metropolis rule,
the buffer size $L$ at each node equals to $100$
and we only take into consideration information that is coming
from the single hop neighbors.
Finally, the adopted performance metric is the average MSE, with definition
$MSE(n)=1/K \sum_{k\in\N} (d_k(n)-f_{k,n}(\x_k(n))$.
As it can be seen from Fig.~\ref{fig:exp1} 
the KDLMS outperforms the other LMS variants, since it
converges faster to a lower error floor compared to them.
In the second experiment the setup is similar to the previous one
albeit here we increase the variance  of the noise, which now equals to  $10^{-1}$.
Fig.~\ref{fig:exp2} illustrates that the enhanced performance of the KDLMS, compared to the other algorithms, is retained in this scenario as well.
\vspace{-0.5cm}
\section{Conclusions and Future Research}
In this paper,
 a novel Kernel based Diffusion LMS, suitable for
 non--linear distributed adaptive filtering was proposed.
  	The theoretical properties of the algorithm were discussed
    and the performance  of the scheme  was tested against
    other adaptive strategies.
 Future research focuses on accelerating the convergence 
 speed by utilizing more data per iteration, 
 as well as investigating sophisticated strategies, which 
 reduce the number of coefficients   by storing the
 most ``informative" ones. 


%
%
\bibliographystyle{IEEEbib}
\bibliography{strings,refs}

\begin{thebibliography}{10}

\bibitem{SlGiMa14}
Konstantinos Slavakis, Georgios Giannakis, and Gonzalo Mateos,
\newblock ``Modeling and optimization for big data analytics:(statistical)
  learning tools for our era of data deluge,''
\newblock {\em IEEE Signal Processing Magazine}, vol. 31, no. 5, pp. 18--31,
  2014.

\bibitem{ChSaLi07}
Cheng Chu, Sang~Kyun Kim, Yi-An Lin, YuanYuan Yu, Gary Bradski, Andrew~Y Ng,
  and Kunle Olukotun,
\newblock ``Map-reduce for machine learning on multicore,''
\newblock {\em Advances in neural information processing systems}, vol. 19, pp.
  281, 2007.

\bibitem{zikopoulos2011understanding}
Paul Zikopoulos, Chris Eaton, et~al.,
\newblock {\em Understanding big data: Analytics for enterprise class hadoop
  and streaming data},
\newblock McGraw-Hill Osborne Media, 2011.

\bibitem{Sa13}
Ali~H Sayed,
\newblock ``Diffusion adaptation over networks,''
\newblock {\em Academic Press Library in Signal Processing}, vol. 3, pp.
  323--454, 2013.

\bibitem{ChSlTh12}
Symeon Chouvardas, Konstantinos Slavakis, and Sergios Theodoridis,
\newblock ``Adaptive robust distributed learning in diffusion sensor
  networks,''
\newblock {\em IEEE Transactions on Signal Processing}, vol. 59, no. 10, pp.
  4692--4707, 2011.

\bibitem{LoSa08}
Cassio~G Lopes and Ali~H Sayed,
\newblock ``Diffusion least-mean squares over adaptive networks: Formulation
  and performance analysis,''
\newblock {\em IEEE Transactions on Signal Processing}, vol. 56, no. 7, pp.
  3122--3136, 2008.

\bibitem{CaSa10}
Federico~S Cattivelli and Ali~H Sayed,
\newblock ``Diffusion {LMS} strategies for distributed estimation,''
\newblock {\em IEEE Transactions on Signal Processing}, vol. 58, no. 3, pp.
  1035--1048, 2010.

\bibitem{ScMaGi09}
Ioannis~D Schizas, Gonzalo Mateos, and Georgios~B Giannakis,
\newblock ``Distributed {LMS} for consensus-based in-network adaptive
  processing,''
\newblock {\em IEEE Transactions on Signal Processing}, vol. 57, no. 6, pp.
  2365--2382, 2009.

\bibitem{MaScGi09}
Gonzalo Mateos, Ioannis~D Schizas, and Georgios~B Giannakis,
\newblock ``Distributed recursive least-squares for consensus-based in-network
  adaptive estimation,''
\newblock {\em Signal Processing, IEEE Transactions on}, vol. 57, no. 11, pp.
  4583--4588, 2009.

\bibitem{ChCeSa14}
Jie Chen, C{\'e}dric Richard, and Ali~H Sayed,
\newblock ``Multitask diffusion adaptation over networks,''
\newblock {\em IEEE Transactions on Signal Processing}, vol. 62, no. 16, pp.
  4129--4144, 2014.

\bibitem{PlBoBe15}
Jorge Plata-Chaves, Nikola Bogdanovic, and Kostas Berberidis,
\newblock ``Distributed diffusion-based {LMS} for node-specific adaptive
  parameter estimation,''
\newblock {\em IEEE Transactions on Signal Processing}, vol. 63, no. 13, pp.
  3448--3460, 2015.

\bibitem{bouboulis2011extension}
Pantelis Bouboulis and Sergios Theodoridis,
\newblock ``Extension of wirtinger's calculus to reproducing kernel hilbert
  spaces and the complex kernel {LMS},''
\newblock {\em IEEE Transactions on Signal Processing}, vol. 59, no. 3, pp.
  964--978, 2011.

\bibitem{bouboulis2012augmented}
Pantelis Bouboulis, Sergios Theodoridis, and Michael Mavroforakis,
\newblock ``The augmented complex kernel {LMS},''
\newblock {\em IEEE Transactions on Signal Processing}, vol. 60, no. 9, pp.
  4962--4967, 2012.

\bibitem{slavakis2008online}
Konstantinos Slavakis, Sergios Theodoridis, and Isao Yamada,
\newblock ``Online kernel-based classification using adaptive projection
  algorithms,''
\newblock {\em IEEE Transactions on Signal Processing}, vol. 56, no. 7, pp.
  2781--2796, 2008.

\bibitem{slavakis2009adaptive}
Konstantinos Slavakis, Sergios Theodoridis, and Isao Yamada,
\newblock ``Adaptive constrained learning in reproducing kernel hilbert spaces:
  the robust beamforming case,''
\newblock {\em IEEE Transactions on Signal Processing}, vol. 57, no. 12, pp.
  4744--4764, 2009.

\bibitem{DKLMSrichards}
Wei Gao, Jie Chen, C{\'e}dric Richard, and Jianguo Huang,
\newblock ``Diffusion adaptation over networks with kernel least-mean-square,''
\newblock {\em Computational Advances in Multi-Sensor Adaptive Processing
  (CAMSAP), 2015 IEEE International Workshop on, 2015 (submitted).}

\bibitem{Ha08}
Simon~S Haykin,
\newblock {\em Adaptive filter theory},
\newblock Pearson Education India, 2008.

\bibitem{SmScBe98}
Alex~J Smola and Bernhard Sch{\"o}lkopf,
\newblock {\em Learning with kernels},
\newblock Citeseer, 1998.

\bibitem{Th15}
Sergios Theodoridis,
\newblock {\em Machine Learning: A Bayesian and Optimization Perspective},
\newblock Academic Press, 2015.

\bibitem{LiPoPuPr08}
Weifeng Liu, Puskal~P Pokharel, and Jose~C Principe,
\newblock ``The kernel least-mean-square algorithm,''
\newblock {\em IEEE Transactions on Signal Processing}, vol. 56, no. 2, pp.
  543--554, 2008.

\bibitem{GaChRiHu14}
Wei Gao, Jie Chen, Cedric Richard, and Jianguo Huang,
\newblock ``Online dictionary learning for kernel {LMS},''
\newblock {\em IEEE Transactions on Signal Processing}, vol. 62, no. 11, pp.
  2765--2777, 2014.

\bibitem{NaPa90}
Kumpati~S Narendra and Kannan Parthasarathy,
\newblock ``Identification and control of dynamical systems using neural
  networks,''
\newblock {\em IEEE Transactions on Neural Networks}, vol. 1, no. 1, pp. 4--27,
  1990.

\bibitem{Ma04}
Danilo~P Mandic,
\newblock ``A generalized normalized gradient descent algorithm,''
\newblock {\em IEEE Signal Processing Letters}, vol. 11, no. 2, pp. 115--118,
  2004.

\end{thebibliography}

\end{document}